# The mechanisms of hot salt stress corrosion cracking in titanium alloy Ti-6Al-2Sn-4Zr-6Mo

Sudha Joseph[a], Trevor C Lindley[a], David Dye[a], Edward A Saunders[b]

[a]Department of Materials, Royal School of Mines, Imperial College London, Prince Consort Road, London SW7 2BP, UK

[b]Rolls-Royce plc., Materials - Failure Investigation, Bristol BS34 7QE, UK

**Abstract**

Hot salt stress corrosion cracking in Ti 6246 alloy has been investigated to elucidate the chemical mechanisms that occur. Cracking was found to initiate beneath salt particles in the presence of oxidation. The observed transgranular fracture was suspected to be due to hydrogen charging; XRD and high-resolution transmission electron microscopy detected the presence of hydrides that were precipitated on cooling. SEM-EDS showed oxygen enrichment near salt particles, alongside chlorine and sodium. Aluminium and zirconium were also involved in the oxidation reactions. The role of intermediate corrosion products such as $Na_2TiO_3$, $Al_2O_3$, $ZrO_2$, $TiCl_2$ and TiH are discussed.

## 1. Introduction

Titanium alloys have good corrosion resistance in most circumstances, which makes them attractive in a variety of applications in aerospace and chemical industries [1]. However, these alloys are susceptible to hot salt stress corrosion cracking (HSSCC) when exposed to halides in the temperature range 200-500°C [2-7]. HSSCC was first reported in Ti-6Al-4V alloy that was creep tested at 370°C [8]. It has been recently shown that fatigue cracking developed from hot salt stress corrosion is characterised with a distinctive blue spot in low cycle fatigued Ti 6246 alloy [9, 10]. Similar HSSCC has also been observed in hot cyclic fatigue spin tests carried out on rotating Ti 6246 gas turbine components [11]. HSSCC in titanium alloys occurs as a two-stage process: crack initiation driven by surface chemistry, followed by crack propagation dependent upon the corrosion kinetics, fracture mechanics and material properties. In general, it is believed that atomic hydrogen produced during the electrochemical reaction between titanium and salt at elevated temperatures is actually responsible for the failure [2,7,12-14]. However, the mechanism by which hydrogen embrittles titanium alloys at high temperatures is not yet fully understood. Hydrogen embrittlement has been suggested to occur by hydrogen decohesion [15], hydrogen enhanced localized plasticity (HELP) [9,10,16,17] and/or brittle titanium hydride formation [7]. In addition to HELP, Absorption Induced Dislocation Emission (AIDE) has recently been observed during aqueous stress corrosion cracking by S Cao et al.[18].

This kind of hot salt stress corrosion failure is mainly observed in laboratory samples and not in aircraft components even though the aircraft environment can often be contaminated by halides. This could be because the airflow in flight might remove corrosion products, preventing the reactions leading to embrittlement. Alternatively, it has been suggested [9, 11] that at high pressures (e.g. in a compressor) an oxidation reaction is favoured over the low pressure hydrolysis reaction with moisture [8, 12], suppressing hydrogen charging and the corresponding drop in fracture toughness. Therefore the reactions occurring at high pressures will be different from those at low pressures in the laboratory [1, 19]. It has been reported that numerous reactions are possible in titanium alloys and number of models have been proposed by various authors to describe the mechanisms of failure [3, 20, 21].

The main concern of the present work is to investigate the various corrosion products formed during testing, and thereby understand the chemical reactions responsible for failure. In previous work the intermediate corrosion products have not been observed directly; confirmation of the reaction sequence is desirable, as it will improve confidence that these are correctly understood, allowing the integrity of service condition to be established. In this study the jet engine compressor disc alloy Ti-6Al-2Sn-4Zr-6Mo is examined using detailed conventional and analytical TEM in order to make such observations of the reaction sequence.

## 2. Material and experimental procedure

Rectilinear 60 x 3.5 x 1.5 mm samples were removed from a Ti 6246 forging that was received in a representative condition, Fig. 1a. A 2-point bend rig [22] was used for loading the sample within a furnace, Fig. 1b. Tests were carried out at 350 and 450°C for different durations under a surface stress of 520MPa, below the yield point of the alloy. The test matrix is shown in Table 1. The samples were polished and a droplet of NaCl salt solution (0.1 g of NaCl in 100ml deionized water) was applied at the center of each sample and allowed to dry before loading the sample in the bend rig. The required stress was achieved by adjusting the holder span of the rig, H

$$H = \frac{KtE(T)\sin\left(\frac{L\sigma}{KtE(T)}\right)}{\sigma} \qquad (1)$$

where H is the distance between the two ends of the holder to achieve the desired stress σ, L and t are the length and thickness of the specimen, E(T) is the Young's modulus of the alloy at temperature T, and K is an empirical constant equal to 1.28. The true thickness of the sample was measured prior to loading, after polishing. Both the dimensions of the sample and H were measured using a calibrated digital vernier caliper with an accuracy of $\pm$ 0.1 mm. At the test temperatures considered here, a stress drop of approximately 5MPa is expected due to differential thermal expansion of the steel rig with respect to the sample.

Phase identification of the corrosion products on the sample surface were carried out using a Bruker X-ray diffractometer (D2 Phaser) with Cu Kα radiation of wavelength 1.54Å. A Zeiss Auriga field emission gun scanning electron microscope (FEG-SEM) in secondary electron imaging mode was used for metallography and fractography. Electron backscattered (BSE) imaging was used for phase analysis. The chemical compositions of the corrosion products near to the salt particles were measured using energy dispersive X-ray spectroscopy (EDS). For this analysis, an accelerating voltage of 10kV and a 60μm aperture were used throughout. To view the corrosion products beneath the salt particles, a cross section of the sample was made in the same SEM by $Ga^+$ focused ion beam milling.

Transmission electron microscopy (TEM) and scanning TEM (STEM) analysis of the corrosion products were carried out on a failed sample using a JEOL JEM-2100F TEM/STEM operated at 200kV. The microscope was equipped with an Oxford Instruments X-Max 80 $mm^2$ silicon drift detector for chemical analysis. The TEM samples for this analysis were prepared using the focused ion beam (FIB) lift-out technique in a dual beam FEI Helios NanoLab 600 using a 30kV $Ga^+$ ion beam. TEM foils were lifted out from both the top surface and fracture surface of the failed specimen. To protect the area of interest, a gas injection system was used to deposit a platinum-containing protective layer. Samples were then made electron transparent by thinning down to a thickness of 150nm. EDS chemical analysis was carried out in STEM mode using a 1 nm probe.

## 3. Results

### 3.1 Microstructure

The alloy was β forged, solution treated in α+β phase field, fan air cooled from just below the β transus and then aged at 600°C. This process gives coarse primary α ($α_p$) and fine scale secondary α ($α_s$) in β. The BSE image in Fig. 2 shows the coarse and elongated $α_p$ as well as much finer $α_s$ laths residing within a retained β phase matrix.

### 3.2 Corrosion test results

#### 3.2.1. Observations on the 350°C test:

Tests were carried out for 24, 60 and 350 hours. The 24h exposure did not show any oxidation near to the salt particles, Fig. 3a. The 60h exposure was found to have a blue coloration, Fig. 3b, and corrosion deposits around the salt particles, Fig. 3c. Only a few of the many salt particles were observed to have corrosion products nearby and these were only in small quantities. Oxidation was found to increase with time. In this context, oxidation is taken to indicate the formation of oxide corrosion products by pyro-hydrolysis in the presence of salt at elevated temperatures. The 350h sample was periodically examined during testing and no cracks were observed up to 250h. A single crack was observed after a long exposure of 350h, which initiated from a salt particle where significant oxidation had occurred, Fig. 3d. The incubation period for crack initiation at this temperature was therefore between 250 and 350h.

#### 3.2.2. Observations on the 450°C test:

Significant blue coloration, Fig. 4a, and oxidation near salt particles were observed after 24h of exposure. But no cracks observed. The sample failed after 60h exposure. The cracks were initiated beneath salt particles where oxidation had occurred, Fig. 4b and the sample failed by the coalescence of cracks, Figs. 4c and d.

#### 3.2.3. SEM-EDS quantification:

The chemical compositions of the corrosion products near the salt particles were obtained by SEM-EDS, as shown in Table 2. The mean values from 5 spatial locations are provided. All the tests showed oxidation products associated with the salt particles, except for the 350°C/24h test. In general, oxidation increased with test temperature and time. Al was found to be involved in the oxidation reactions. Na and Cl were observed in the corrosion products and the Na level in the corrosion products was smaller in the 350°C test sample when compared to the 450°C sample.

#### 3.2.4. X-ray diffraction analysis:

The different phases formed during the hot salt stress corrosion tests were examined by X-ray diffraction (XRD). The XRD patterns were obtained from the corrosion deposits on top of the failed 450°C/60h sample, Fig. 5. For comparison, the XRD pattern of the as-received sample is also shown. Due to various alloying elements and numerous chemical reactions, a large number of phases are expected. New peaks were observed at diffraction angles of 27.5°, 29.5°, 31.8°, 35.8°, 45.5° and 66° after the test. Out of these six peaks, the two peaks at 29.5° and 35.8° are unique and not overlapping with the peaks for any other phases. The peak at 29.5° corresponds to monoclinic sodium titanate $Na_2TiO_3$ (JCPDS card 37-0345). The peak at 35.8° is shown more clearly in the inset of the figure. This peak corresponds to the 100% intensity peak of titanium hydride (JCPDS card 40-1244). The other four peaks match with oxides such as $Al_2O_3$ (JCPDS card 42-1468), $ZrO_2$ (JCPDS card 37-1484) and the alloy chloride, $TiCl_2$ (JCPDS card 77-0188). The intensity of many of the $TiO_2$ peaks decreased after the test, implying that the naturally occurring passive oxide layer was consumed during the corrosion process. Thus the XRD analysis indicates the formation of sodium titanate and titanium hydride, and the consumption of the original passive titanium oxide film.

### 3. 2. 5 Cross-section:

A cross sectional view of the 450°C/60h tested sample is shown in Fig. 6. The FIB cross-section was made at the edge of a salt particle, Fig. 6a, so that the oxidation products could be observed in the presence and absence of the salt. The cross section of the region is shown in Fig. 6b. A single oxide layer was observed on the base alloy in the "no salt" region, Fig. 6c. This oxide layer was found to contain a small amount of Al, suggesting that Al starts diffusing first during the test. In the salt region three different kinds of layered deposits on the base alloy were observed, marked by numbers in Fig. 6d. Table 3 shows the composition of these various oxide layers on the top of the base alloy, with each measurement being the average of five measurement locations. These are titanium oxide with a small amount of Al in layer 1, Al and Zr in layer 2 and Al, Zr and Mo in layer 3. There was a gradual increase of Al from the bottom layer 3 to the top most layer 1. It was also noticed that the Zr and Mo content is zero at the first layer and increases across the layers, presumably due to the difference in mobility between Al, Zr and Mo. Hence there is an outward diffusion of alloying elements from the base alloy, which would create vacancies. Layer 2a at the right edge possessed a darker contrast to layer 2, which is the reaction product of layer 2 with the NaCl salt. It was also observed that the top most region (around 0.5μm depth) of the base alloy was enriched in oxygen, which is shown by a dashed line. The composition of this oxidized region is shown in Table 3d. The oxidation levels were higher where there was a reaction with salt particles. There were also several secondary cracks in this region.

### 3. 2. 6. TEM analysis:

TEM analysis was performed on FIB lift-out foils from the top surface and the fracture surface of the failed 450°C/60h sample. Fig. 7a shows a BF-STEM image of the sample from the fracture surface, comprising a corrosion deposit on top of an oxidized layer of ~1μm depth. A STEM-EDS line scan is shown in Fig. 7c, where the scan was taken along the line shown in Fig. 7b. The micrograph in Fig. 7b gives a magnified view of the region marked in Fig. 7a. Na, Al and Zr were observed in the corrosion deposit, along with an O content greater than in the oxidized layer. Elevated Cl levels were observed at the alloy/oxide interface, which might cause metal/oxide debonding. Selected area diffraction (SAD) patterns from the different layers of the sample are shown in Fig. 7 d & e. The ring pattern obtained from the corrosion products is due to the nanoscale dimensions of the precipitates. The d-spacings of the rings matched with $TiO_2$, $Al_2O_3$, $ZrO_2$, $TiCl_2$ and TiH. The oxidized metal possessed a $TiO_2$ single crystal SAD pattern. The HRTEM images of the corrosion layer are shown in Fig. 8. A variety of precipitates with different d-spacings were observed in the image. Their d-spacings match those of $ZrO_2$, $TiCl_2$ and TiH phases. The inserts in the figures are fast Fourier Transforms (FFT) of the images corresponding to the diffraction of various precipitates and titanium oxide. It should be noted that the solubility for H will be much higher at the test temperature [23] and therefore it is likely that TiH formation occurred on cooling. Na and Cl were detected in the metal below the oxidized layer, Fig. 9. Sodium titanate formation was also observed, Fig. 10a. The EDS spectrum and composition of the product are shown in Fig. 10 b. A ring pattern (Fig. 10c) was obtained from these deposits, suggesting that these titanates formed as nanosized precipitates, Fig. 10 d. In addition to sodium titanate, the formation of zirconium oxychloride was observed on the foil taken from the top surface of the sample, Fig. 11. This figure shows the DF-STEM image of the region and a corresponding STEM-EDS map. Shear of a β plate was also observed, Fig. 12a. The EDS spectrum obtained from the corrosion products associated with the shear of the β plate is shown in Fig. 12b. The Cu peak in Fig. 10b and 12b came from the TEM sample holder and was not considered during quantification.

### 3. 2. 7 Fractography:

The fracture surface of the 450°C/60h sample in Fig. 13a possessed a brittle and microstructure-sensitive appearance. The growth of secondary cracks was also observed just below the salt particle. The higher magnification image in Fig. 13b shows the corrosion deposits on the fracture surface. In Fig. 13c transgranular fracture was observed in the stress corrosion cracking region, suspected to be due to hydrogen charging from stress corrosion. In the final failure region the alloy was observed to have failed by ductile microvoid coalescence, Fig. 13d.

## 4. Discussion:

Ti alloys are often found to be more susceptible to HSSCC than commercial purity Ti [4, 24-26], which might imply that the alloying additions play a significant role in the chemical reactions that occur during the process. The schematic in Fig. 14 depicts the overall mechanisms involved in this hot salt stress corrosion failure. The sample initially has a protective oxide film upon which a salt deposit is applied. Under the applied stress and temperature, different layers of corrosion deposits form on top of the base alloy, as mentioned in Sec. 3.2.5. Al was found to diffuse in the oxide layer where there is no salt paticles, forming layer 1. In the presence of salt, Zr and Mo also diffused outward in addition to Al, leading to the formation of further layers of corrosion deposit, layers 2 and 3. This sequence follows the diffusivity of each of these alloying elements in the phases of the base metal. Thus, the salt enhances the diffusion of number of alloying elements, promoting various chemical reactions. High resolution imaging and selected area diffraction by TEM on the corrosion layers above the oxidized layer found different oxides such as $Al_2O_3$, $ZrO_2$ and zirconium oxychlorides as end products. They were found in the form of nanoprecipitates. XRD patterns supported the presence of these oxides within the corrosion layers. This is in agreement with previous results that the susceptibility to HSSCC increases with Al content in the alloy [24] and that Al is preferentially attacked by NaCl [20, 27]. At the same time, oxygen ions diffuse inward through the metal/oxide interface up to 1μm into the base alloy resulting in the oxidation of the base alloy. There is also speculation that Zr and Mo are protective [19]. Zirconium oxychloride appeared as a result of the reactions between NaCl and $ZrO_2$. Zirconium oxychlorides have previously been assumed to be inactive in the corrosion process, despite the fact that they are observed, and have been left out of the simplified reaction sequence [28, 29]. Sn was not observed in the corrosion products suggesting that gaseous tin chlorides might have formed and escaped from the sample. In addition, chlorine was observed at the alloy/oxide interface (Fig. 7c), which might cause interface debonding. Trapped Cl at this location is an indicator of the hot salt stress corrosion process as identified in [9, 11].

Sodium titanate has been observed by X-ray diffraction and directly from the STEM analysis. $Na_2TiO_3$ can be produced by the following reaction at elevated temperatures T [8]:

$$NaCl_{(s)} + TiO_{2(s)} + H_2O_{(g)} \xrightarrow{T} Na_2TiO_{3(s)} + 2HCl_{(g)} \qquad (2)$$

Thus the salt on the top of the base alloy is likely to react with rutile oxide to produce sodium titanate and gaseous hydrochloric acid, as mentioned in the above reaction. The water needed for the reaction can be available from moisture in the atmosphere, or adsorbed onto the titanium surface [3], or from the salt being hydrous, or water containing [20]. This reaction occurs beneath the salt particles (Fig. 4b), consuming the oxide scale and exposing the underlying titanium base alloy.

As the oxide scale is consumed and not healed, the gaseous hydrochloric acid generated by reaction (2) can access the bare alloy and react with titanium and its alloying elements Al and Zr according to the following chemical reactions.

$$Ti_{(s)} + 2HCl_{(g)} \rightarrow TiCl_{2(s)} + 2H \qquad (3)$$

$$Al_{(s)} + 3HCl_{(g)} \rightarrow AlCl_{3(g)} + 3H \qquad (4)$$

$$Zr_{(s)} + 4HCl_{(g)} \rightarrow ZrCl_{4(g)} + 4H \qquad (5)$$

The alloy chloride melting and sublimation points will also affect the mobility of these alloying elements and the formation of the layered corrosion deposits observed (Sec 3.2.5). The hydrogen produced in the above reactions will then charge the base alloy. Hydride phases might form where the solubility limit is exceeded, including on cool-down of the sample. Such hydride formation was observed in our study, from both XRD and HRTEM analysis. The kinetics of the above reactions increases with temperature and hence the generation of hydrogen increases [6, 19, 30] at higher temperatures. In addition, the diffusion of hydrogen in metals obeys the Arrhenius equation [31] and the diffusion of hydrogen is rapid [32] at higher temperatures. This increasing hydrogen charging is held to be responsible for the failure of the sample at 450°C, whilst notably less cracking was observed in testing at 350°C.

The metal chlorides produced in the above reactions (2)-(4) can undergo the following chemical reactions, resulting in metal oxide formation and the re-generation of HCl.

$$2AlCl_{3(g)} + 3H_2O_{(g)} \rightarrow Al_2O_{3(s)} + 6HCl_{(g)} \qquad (6)$$

$$ZrCl_{4(g)} + 2H_2O_{(g)} \rightarrow ZrO_{2(s)} + 4HCl_{(g)} \qquad (7)$$

These oxides are found as nanoprecipitates in the corrosion deposits. Thus there is a continuous generation of gaseous HCl that reacts with the exposed metal according to equations (2)-(4). This leads to a looping pyrohydrolysis reaction that gives rises to hydrogen charging and thereby cracking of the underlying alloy. These observations are supported by the findings of Rideout et al. [24] and Chevrot [19]. It should be noted that the hydrides observed at room temperature may have precipitated on cool-down of the sample from solution, since the solubility of H in Ti varies substantially with temperature [23].

## 5. Conclusions:

The hot salt stress corrosion cracking of Ti 6246 alloy by NaCl has been studied by X-ray diffraction and electron microscopy in order to elucidate the intermediate chemical reactions involved in the reaction sequence. The following conclusions can be drawn.

- Limited cracking was observed in the alloy at 350°C after a long exposure of 350h. However, fracture occurred at 450°C after 60h of exposure. Cracks were found to initiate beneath the salt particles where oxidation had occurred.
- The alloying elements Al and Zr were found to diffuse outward, dealloying the matrix, and forming layered deposits on the top of the base alloy.
- Elevated Cl levels were observed at the alloy/oxide interface, which can be used as an identifier of the stress corrosion process.
- Various end products such as sodium titanate, alumina, zirconia and titanium chloride are formed and deposited as nanoprecipitates. From these observations, mechanisms for the failure have been proposed.
- The fracture surface was mainly transgranular, brittle in appearance and microstructure-sensitive, supporting the suggestion that hydrogen charging leads to cracking in stress corrosion scenarios. Whilst H in solution cannot be observed directly with the techniques used here, titanium hydrides were identified, which would have precipitated on cooling.
- Some of the β plates were observed to be sheared, associated with the action of the corrosion products.

**Acknowledgements**

Funding from EPSRC under the Hexmat programme grant (EP/K034332/1) is acknowledged.

**References:**

[1]  G. Lütjering, J. C. Williams, Titanium, second ed., Springer, New York, 2003.

[2]  R. K. Dinnappa, Hot salt stress corrosion cracking of a titanium alloy: the phenomenon in view of aero gas turbine operating conditions, Key Eng. Mater. 20–28 (1988) 2255–2271.

[3]  J. R. Myers, J. A. Hall, Hot-salt stress-corrosion cracking of titanium alloys: an improved model for the mechanism, Corrosion 33 (1977) 252–257.

[4]  D. Sinigaglia, G. Taccani, B. Vicentini, Hot salt stress corrosion cracking of titanium alloys, Corros. Sci.18 (1978) 781–796.

[5]  M. Encrenaz, P. Faure, J. A. Petit, Hot salt stress corrosion resistance of Ti 6246 alloy, Corros. Sci. 40 (1998) 939–950.

[6]  M. D. Pustode, V. S. Raja, N. Paulose, The stress-corrosion cracking susceptibility of near-α titanium alloy IMI 834 in presence of hot salt, Corros. Sci. 82 (2014) 191–196.

[7]  M. D. Pustode, V. S. Raja, B. Dewangan, N. Paulose, Effect of long term exposure and hydrogen effects on HSSCC behavior of titanium alloy IMI 834, Mater. Des. 86 (2015) 841-847.

[8]  V. C. Petersen, Hot-salt stress-corrosion of titanium: a review of the problem and methods for improving the resistance of titanium, J. Metals 23 (1971) 40–47

[9]  T. P. Chapman, R. J. Chater, E. A. Saunders, A. M. Walker, T. C. Lindley, D. Dye, Environmentally assisted fatigue crack nucleation in Ti–6Al–2Sn–4Zr–6Mo, Corros. Sci. 96 (2015) 87–101.

[10] T. P. Chapman, V. A. Vorontsov, A. Sankaran, D. Rugg, T. C. Lindley, and D. Dye, The dislocation mechanism of stress corrosion embrittlement in Ti-6Al-

2Sn-4Zr-6Mo, Metall. Mat. Trans. 47A (2016) 282-292

[11] E.A. Saunders, T.P. Chapman, A.R.M. Walker, T.C. Lindley, R.J. Chater, V.A. Vorontsov, D. Rugg, D. Dye, Understanding the "blue spot": Sodium chloride hot salt stress-corrosion cracking in titanium-6246 during fatigue testing at low pressure, Eng. Fail. Anal. 61 (2016) 2-20.

[12] R. S. Ondrejcin, The role of hydrogen in hot-salt stress-corrosion cracking of titanium-aluminium alloy, NASA CR-1915, 1971.

[13] H. R. Gray, Relative susceptibility of titanium alloys to hot-salt stress corrosion, NASA Technical Note TN D-6498, November 1971.

[14] N. G. Plekhanova, E. A. Borisova, V. N. Modestova, T. V. Barysheva, N. D. Tomashov, Salt cracking of titanium alloys, Prot. Met. 12 (1976) 559-561.

[15] H. G. Nelson, A film-rupture model of hydrogen-induced, slow crack growth in acicular alpha-beta titanium, Metall. Trans. 7A (1976) 621–627.

[16] D. S. Shih, I. M. Robertson, H. K. Birnbaum, Hydrogen embrittlement of a titanium: in situ TEM studies, Acta Metall. 36 (1988) 111–124.

[17] S. Cao, C. V. S. Lim, B. Hinton, X. Wu, Effects of microtexture and Ti3Al ($\alpha_2$) precipitates on stress-corrosion cracking properties of a Ti-8Al-1Mo-1V alloy, Corros. Sci. 116 (2017) 22-33.

[18] S. Cao, S. Zhu, C. V. S. Lim, X. Zhou, X. Wu, The mechanism of aqueous stress-corrosion cracking of α + β titanium alloys, Corros. Sci. 125 (2017) 29-39.

[19] T. Chevrot, Pressure Effects on the Hot-Salt Stress-Corrosion Cracking of Titanium Alloys, PhD thesis, School of Industrial and Manufacturing Science, Cranfield University, 1994.

[20] S. P. Rideout, S. P. Louthan, C. L. Selhy, Basic mechanisms of stress-corrosion cracking of titanium, Stress-Corrosion Cracking of Titanium, ASTM STP, 397 (1966) 137-151.

[21] H. L. Logan, Studies of hot-salt cracking of the titanium 8Al-1Mo-1V alloy" Fundamental aspects of stress-corrosion cracking, NACE, Houston. 1969. pp. 662-672.

[22] Haaijer and Loginou, Stress analysis of bent-beam stress corrosion specimens, Corrosion 21 (1965) 105-112.

[23] T.P. Chapman, D. Dye, D. Rugg, Hydrogen in Ti and Zr alloys: industrial perspective, failure modes and mechanistic understanding, Phil. Trans. Roy. Soc. A 375 (2017) 20160418.

[24] P. Rideout, R. S. Ondrejcin, M. R. Louthan Jr, The Science, Technology and Application of Titanium Alloys, Pergamon Press, 1970.

[25] M.J. Donachie Jr, Titanium: A Technical Guide, second ed., ASM International, 2000.

[26] M. Garfinkle, An Electrochemical Model for Hot-Salt Stress-Corrosion of Titanium Alloys, NASA TN D-6779, 1972.

[27] R. S. Ondrejcin, M. R. Louthan Jr., Role of Hydrogen Chloride in Hot-Salt Stress Corrosion Cracking of Titanium–Aluminium Alloys, NASA CR-1133, 1968.

[28] V. C. Petersen, H. B. Bomberger, The mechanism of salt attack on Titanium alloys, in Stress Corrosion Cracking of Titanium, ASTM STP 397 (1966) 80-94.

[29] V. V. Travkin, V. F. Pshirkov, B. A. Kolashev, Thermodynamic analysis of the chemical mechanisms for the hot-salt corrosion of titanium alloys, Sov. Mater. Sci. 15 (1979) 134-137.

[30] G. Martin, Investigation of long term exposure effect under stress of two structural titanium alloys basic mechanisms of stress-corrosion cracking of titanium, ASTM STP 397 (1965) 95-120.

[31] Omar S. Abdul Hamid, Diffusion of Hydrogen in Titanium, Ph.D Thesis, King Fahd University of Petroleum and Mines, 1993.

[32] H. Christ, M. Decker, S. Zeitler, Hydrogen diffusion coefficients in the titanium alloys IMI 834, Ti 10-2-3, Ti 21S and alloy C, Metall. Mater. Trans. 31A (2000) 1507-1517.

# Figure captions

Fig. 1 (a) Ti 6246 alloy sample used for HSSCC testing and (b) 2-point bend rig with a loaded sample.

Fig. 2 BSE image showing the typical microstructure of Ti 6246 alloy comprising primary α and secondary α within a retained β matrix.

Fig. 3 (a) no oxidation near salt particles after 24h exposure (b) blue coloration after 60h exposure (c) oxidation observed near salt particle and (d) crack growth from the salt particle where oxidation occurred. The time of exposure is shown in each micrograph and the test temperature was 350°C.

Fig. 4 (a) severe blue coloration showing oxidation near salt particles, (b) crack initiation beneath the salt particle where oxidation occurred (c) coalescence of cracks leading to failure and (d) photograph showing the failure of the sample. The time of exposure is shown in each micrograph and the test temperature was 450°C.

Fig. 5 X-ray diffraction on the sample tested at 450°C/60h, compared to a pre-test control.

Fig. 6 (a) SEM image of the 450°C/60h sample, showing where the FIB cross section was made, (b) BSE image showing the overall view of the cross section, (c) cross section in the "no salt" region and (d) cross section in the salt region.

Fig. 7 (a) BF-STEM image showing the corrosion deposits on the oxide layer (b) BF-STEM image showing the line scan region (c) STEM-EDS line scan and (d&e) SAD from different layers on the base alloy within the FIB lift out foil from the fracture surface of 450°C/60h sample.

Fig. 8 HRTEM images showing the different precipitates in the corrosion products.

Fig. 9 STEM-BF image showing the deposits on the base alloy below the oxide layer. The composition of the products is also shown.

Fig. 10 (a) STEM dark field image showing the appearance of sodium titanate. (b) STEM-EDS spectrum and EDS quantification showing the composition of the titanate, in at.%. (c) SAD pattern and (d) HRTEM image showing the different orientations of nano-sized sodium titanate precipitates.

Fig. 11 STEM-EDS mapping showing (Zr, Cl, O) deposition in the corrosion product.

Fig. 12 (a) STEM dark field image showing shear of a β plate. (b) EDS spectrum from the corrosion product associated with the shear, and a table showing its composition in at.%.

Fig. 13 (a) Fracture surface showing brittle appearance (b) deposition of corrosion products on the fracture surface (c) transgranular fracture due to hydrogen embrittlement and (d) ductile microvoids. The regions from where the micrographs b, c and d taken are indicated in (a).

Fig. 14 Schematic showing the overall view of the layered corrosion deposits on the base alloy formed during the test.

# Tables

Table 1 Bend test samples (surface stress applied = 520MPa) examined in the present study. *The 250h sample was run on to 350h.

| T(°C) | Duration (hours) | | | |
|---|---|---|---|---|
| 350 | 24 | 60 | 250* | 350 |
| 450 | 24 | 60 | | |

Table 2 SEM-EDS quantification on corrosion products near the salt particles (at. %) under different test conditions.

| Element | 350°C/ 24h | 350°C/ 60h | 350°C/ 350h | 450°C/ 24h | 450°C/ 60h |
|---|---|---|---|---|---|
| Ti | No oxidation products near salt particles | 33.3 | 22.2 | 22.5 | 21.9 |
| Al | | 2.1 | 2.3 | 1.4 | 2.1 |
| Na | | 0.8 | 1.8 | 5.3 | 3.4 |
| Cl | | 1.3 | 1.4 | 6.3 | 1.6 |
| O | | 62.5 | 72.4 | 64.5 | 71.0 |

Table 3 SEM-EDS composition data, in at. %, for the various layered deposits observed in Figure 6 for the 450ºC/60h sample at each location (a-d) and in each layer at those locations.

**a. No salt region**

| Element | Layer 1 |
|---|---|
| Ti | 81.9 |
| O | 11.4 |
| Al | 6.7 |

**b. Salt region – no reaction with salt**

| Element | Layer 1 | Layer 2 | Layer 3 |
|---|---|---|---|
| Ti | 77.0 | 83.1 | 75.6 |
| O | 17.6 | 9.4 | 14.3 |
| Al | 5.4 | 4.4 | 4.5 |
| Zr | 0 | 2.5 | 2.2 |
| Mo | 0 | 0 | 2.9 |

**c. Salt region – reaction with salt**

| Element | Layer 1 | Layer 2a | Layer 3 |
|---|---|---|---|
| Ti | 58.6 | 53.9 | 65.7 |
| O | 19.2 | 9.1 | 16.7 |
| Al | 5.5 | 3.3 | 3.9 |
| Zr | 2.2 | 2.3 | 1.8 |
| Mo | 0 | 1.7 | 2.5 |
| Na | 4.0 | 11.3 | 5.7 |
| Cl | 10.5 | 18.4 | 3.6 |

**d. Base alloy – corrosion affected region**

| Element | No salt region | Salt region - no reaction | Salt region - with reaction |
|---------|---------------|---------------------------|-----------------------------|
| Ti | 81.3 | 81.7 | 66.6 |
| O  | 5.3  | 6.4  | 18.7 |
| Al | 5.5  | 4.5  | 4.1  |
| Zr | 3.4  | 2.1  | 1.8  |
| Mo | 3.4  | 4.3  | 1.9  |
| Sn | 1.1  | 1.1  | 0.72 |
| Na | 0    | 0    | 2.9  |
| Cl | 0    | 0    | 3.2  |

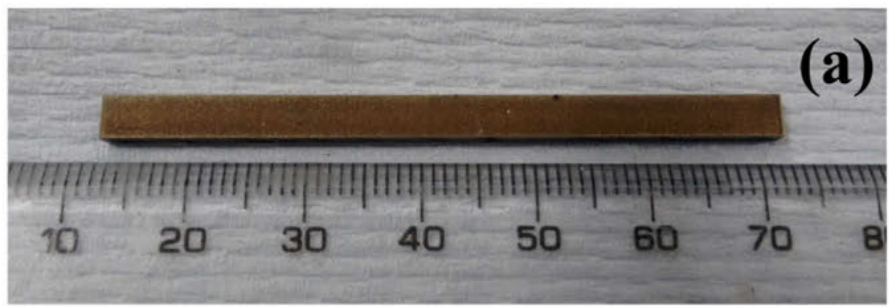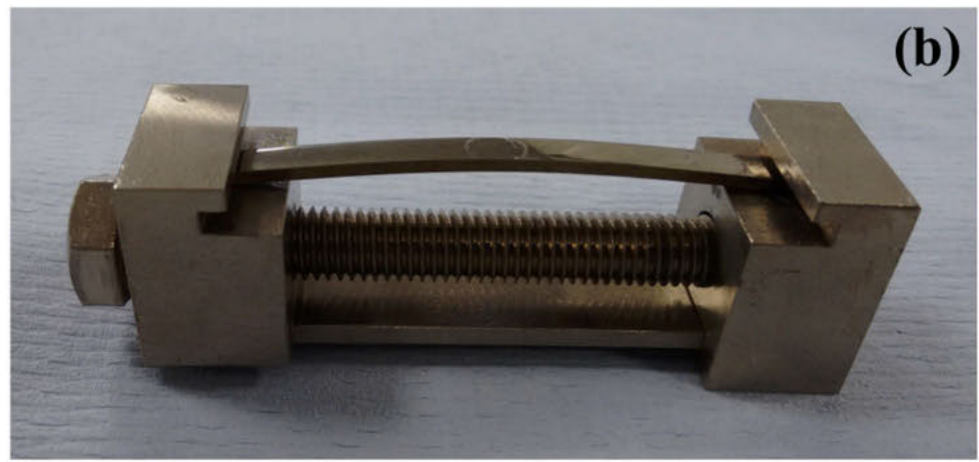

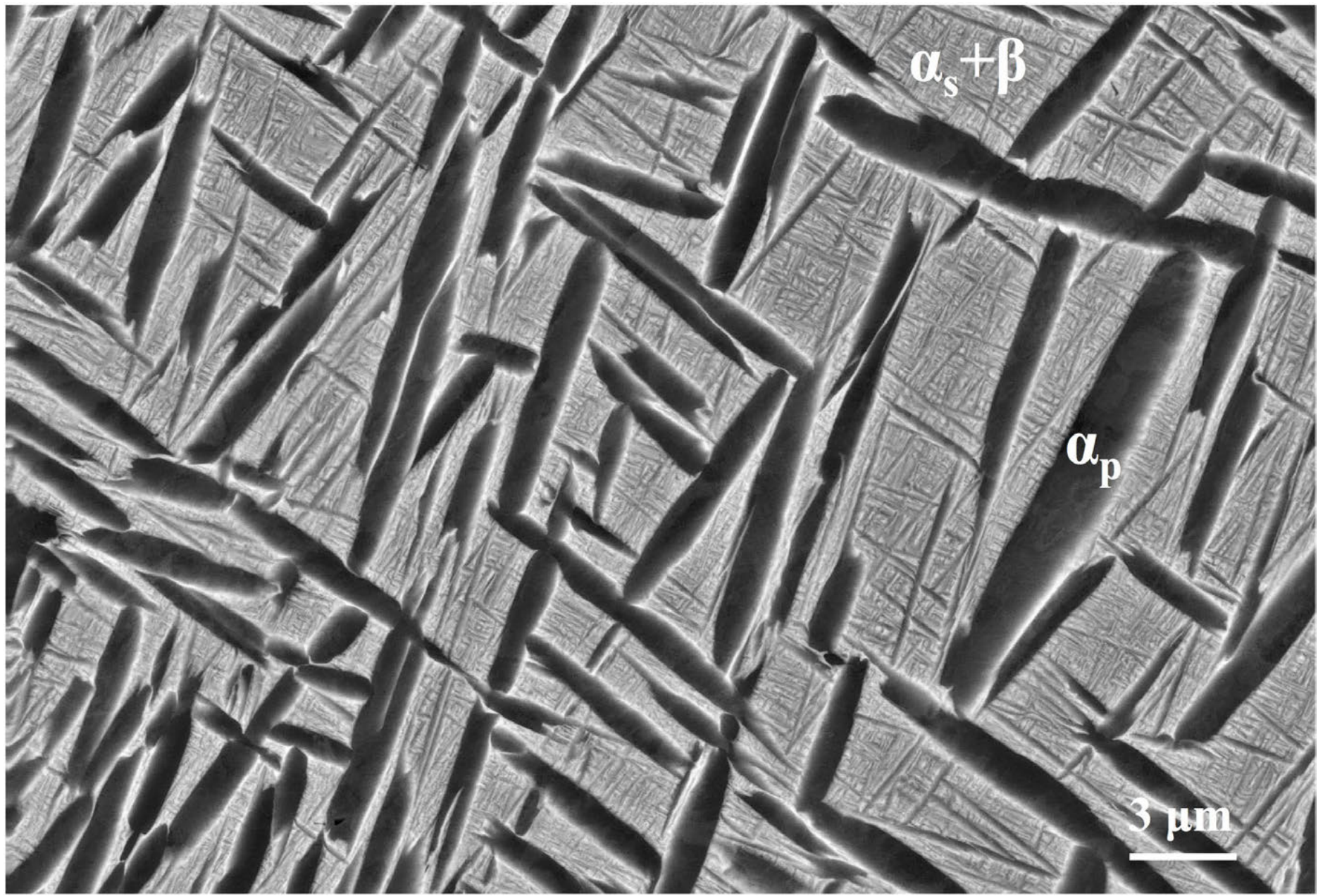

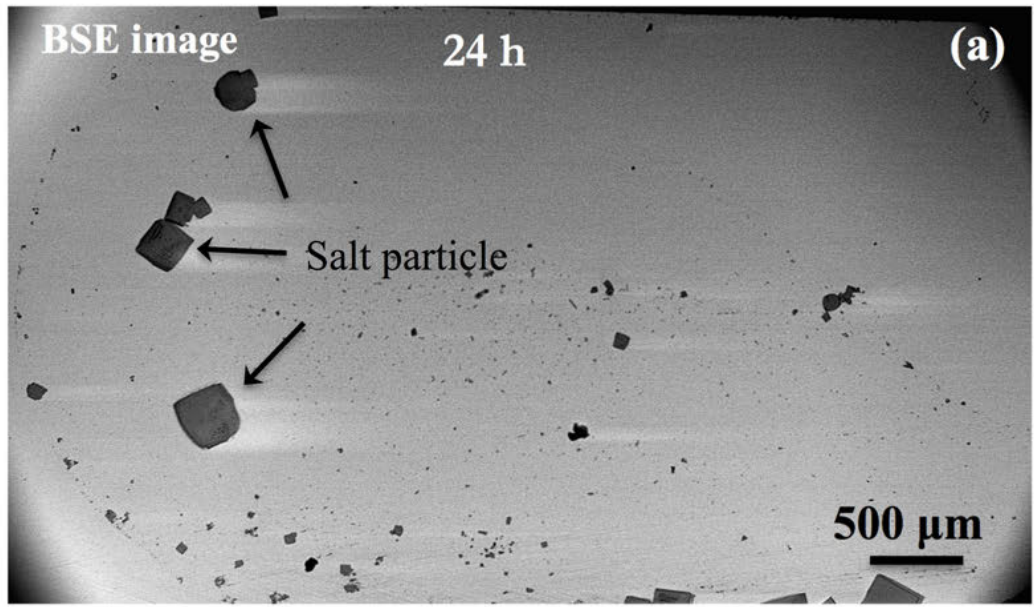
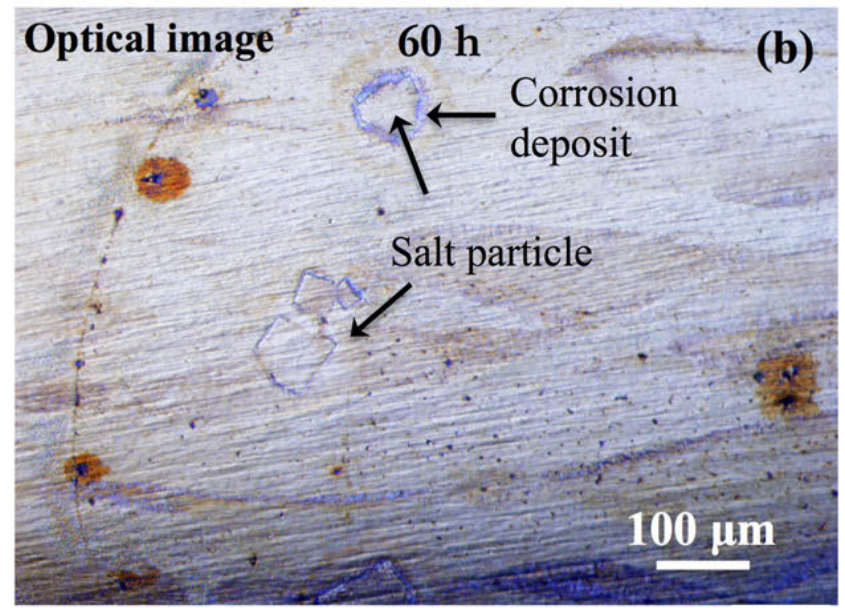
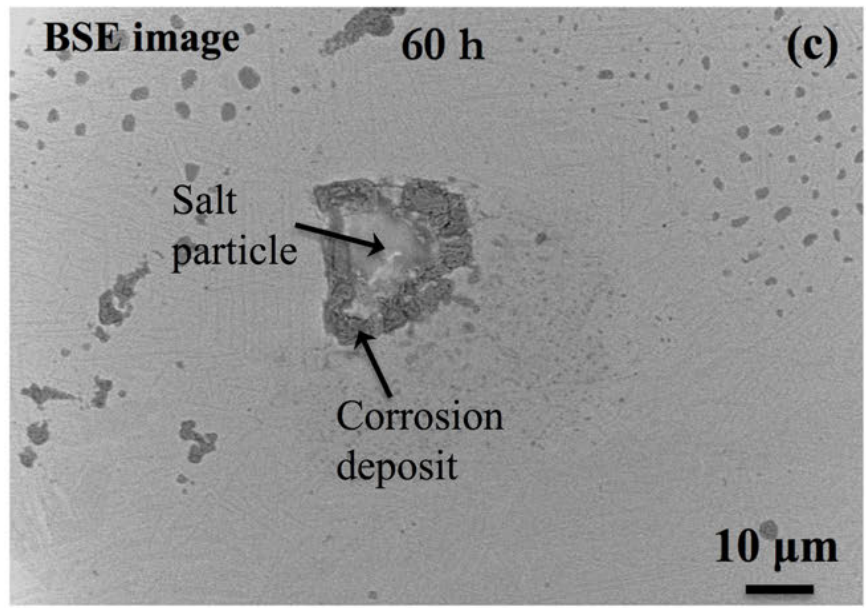
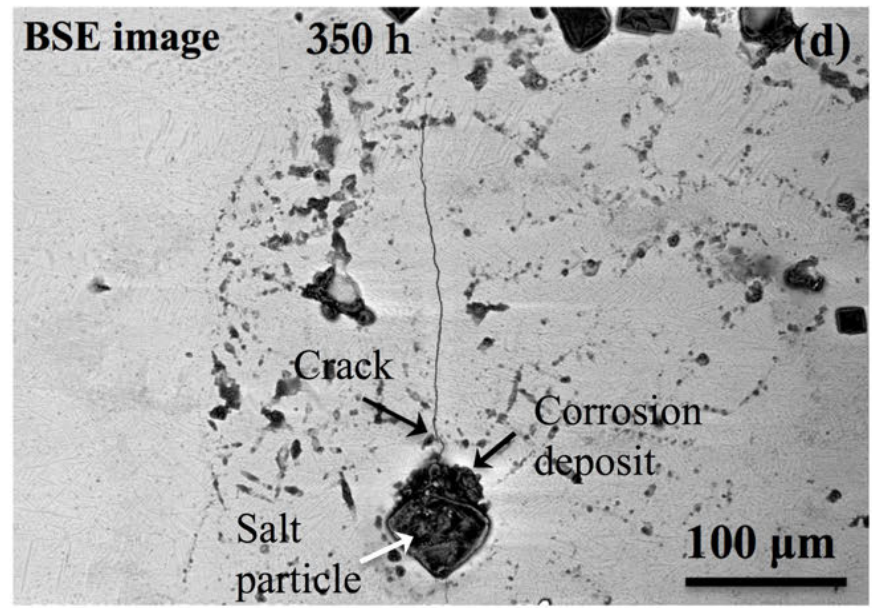

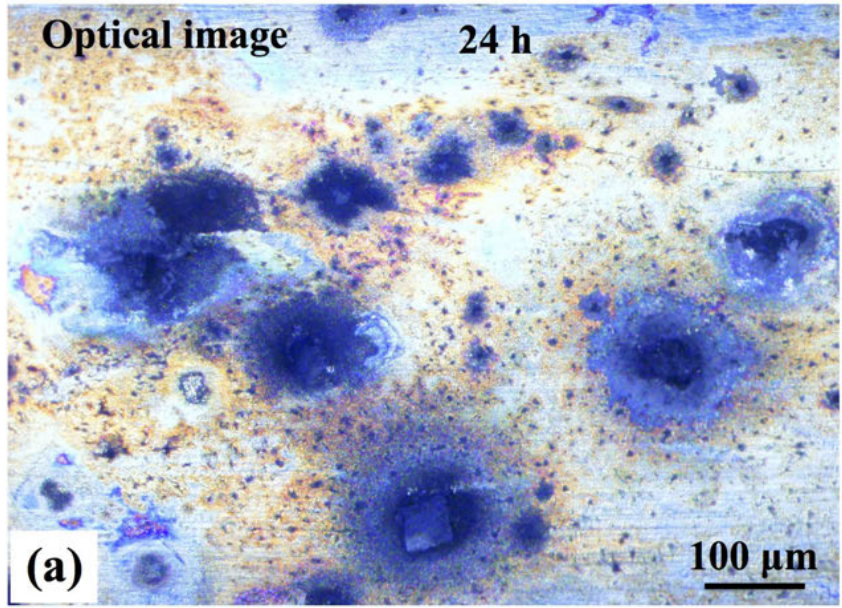 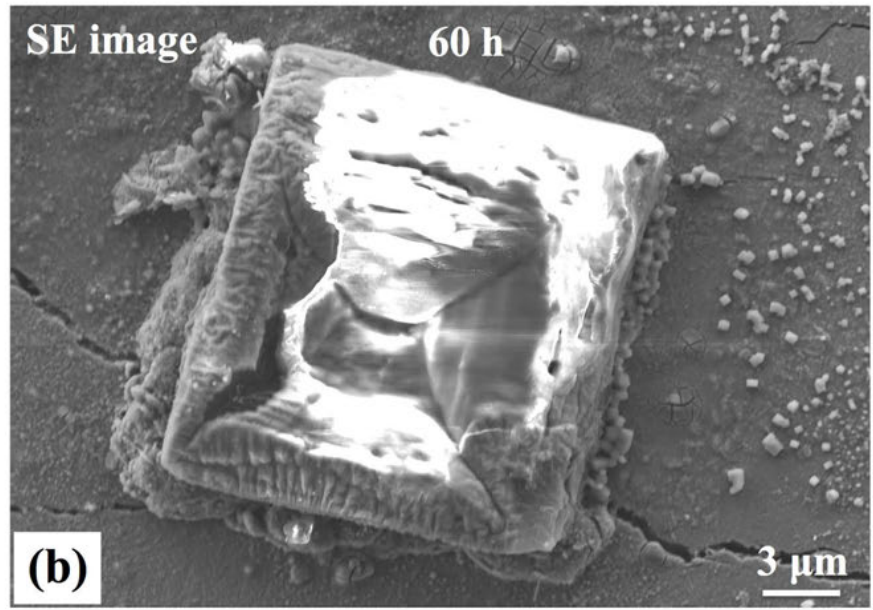
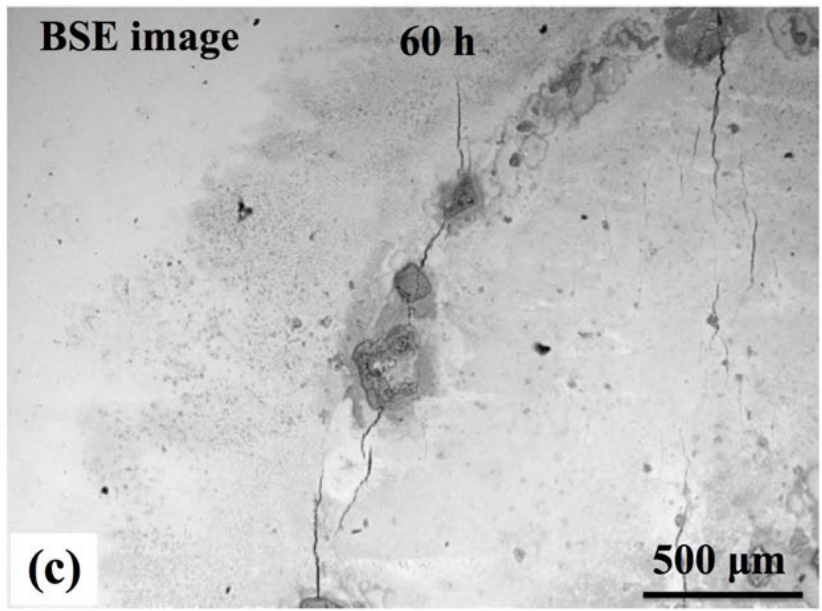 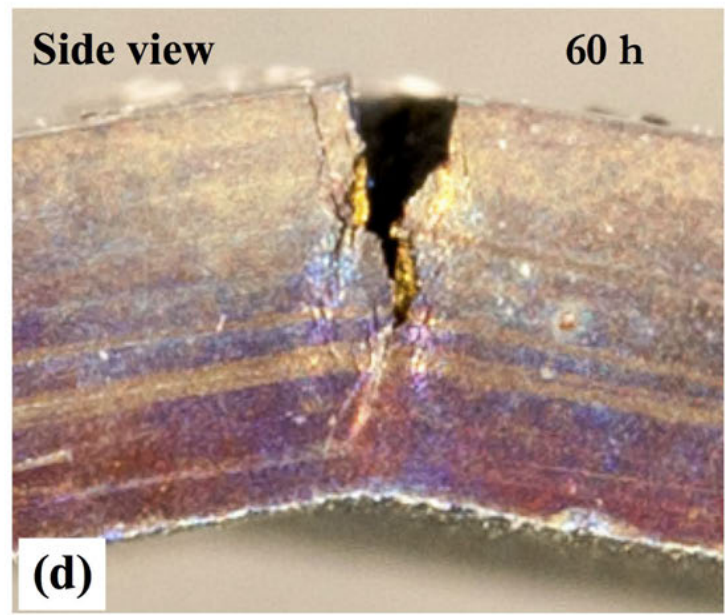

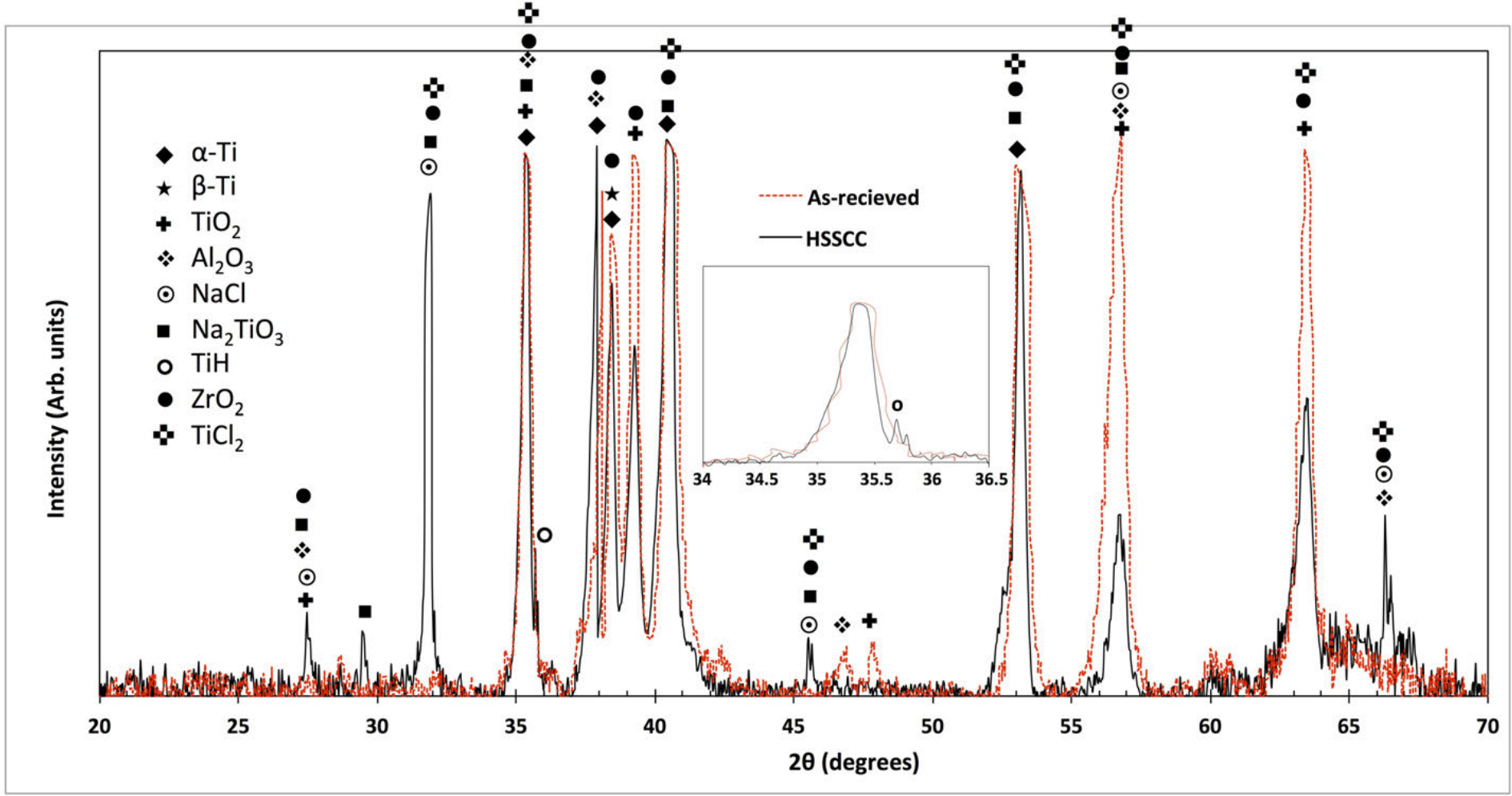

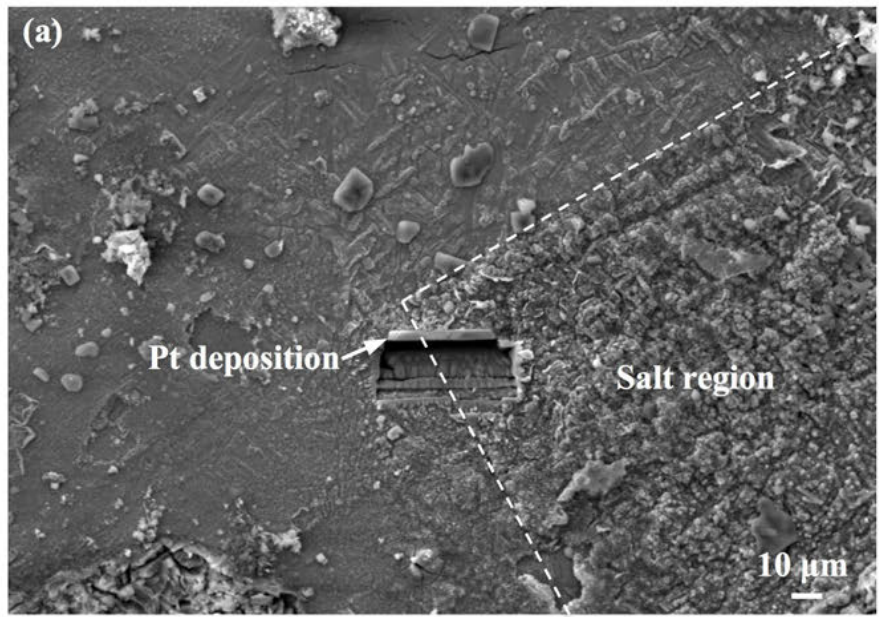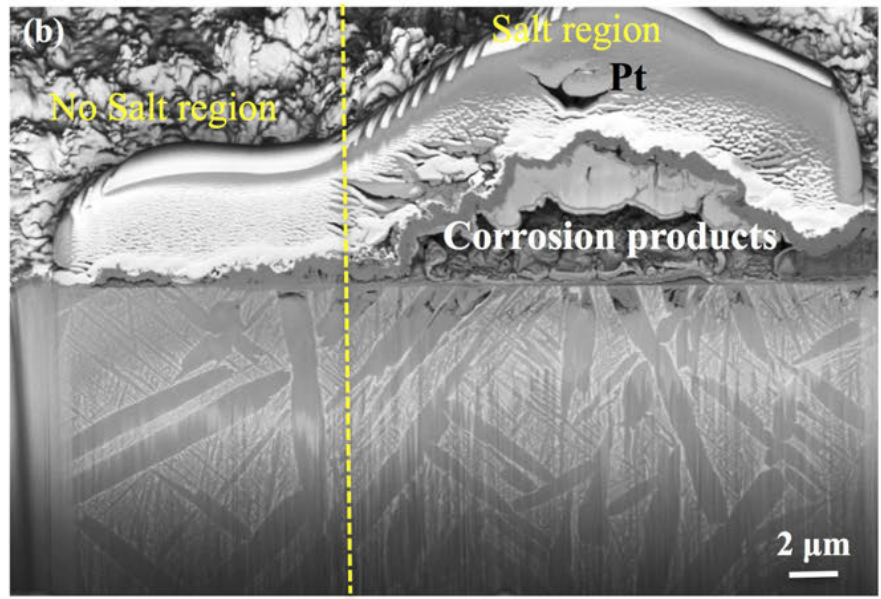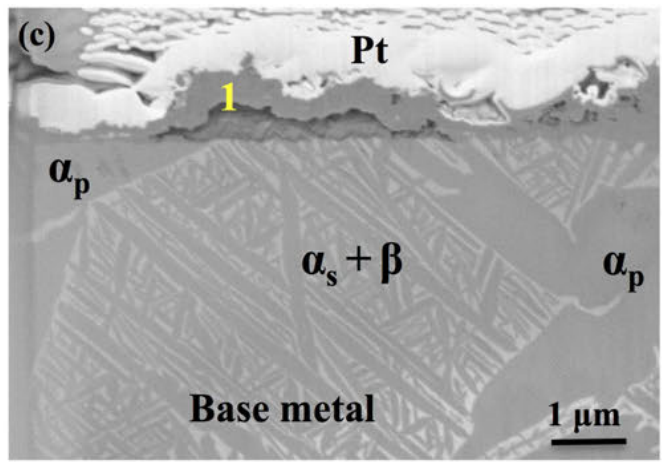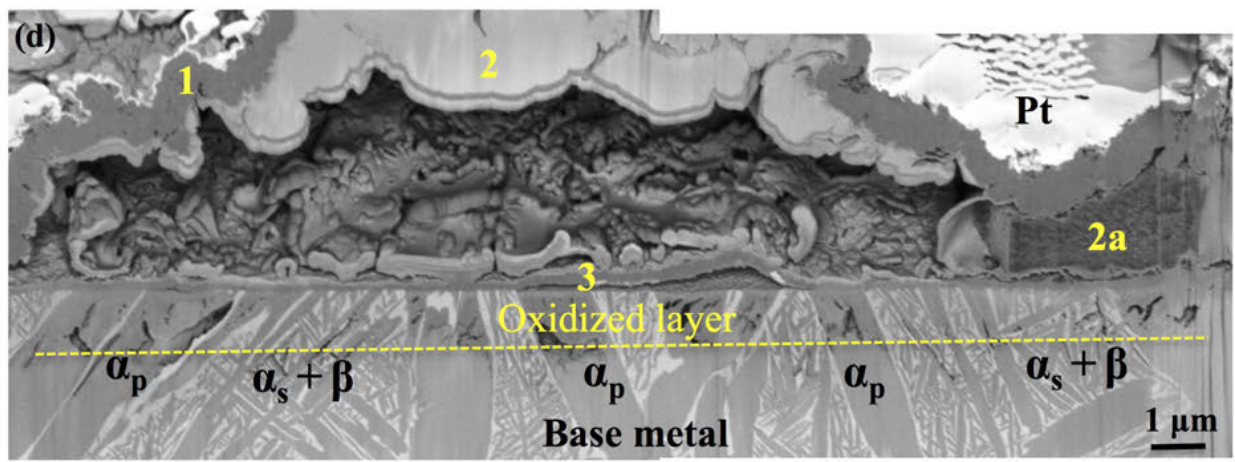

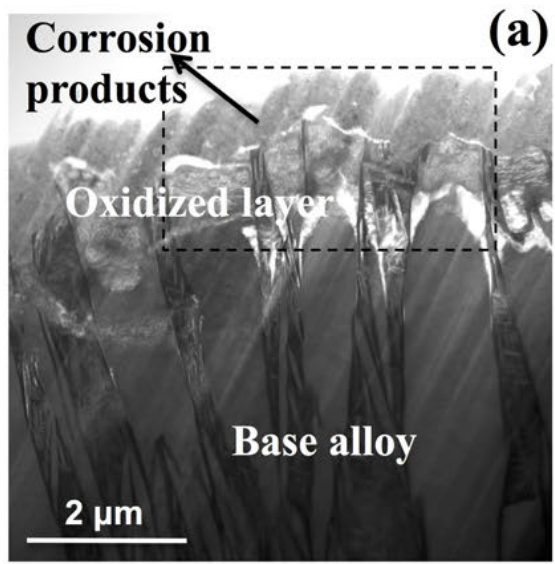 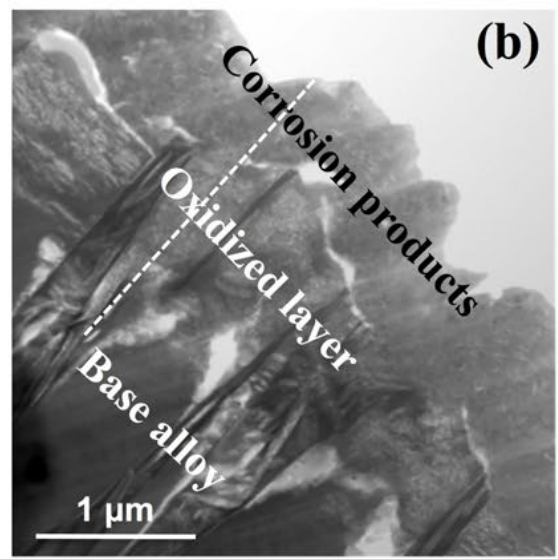 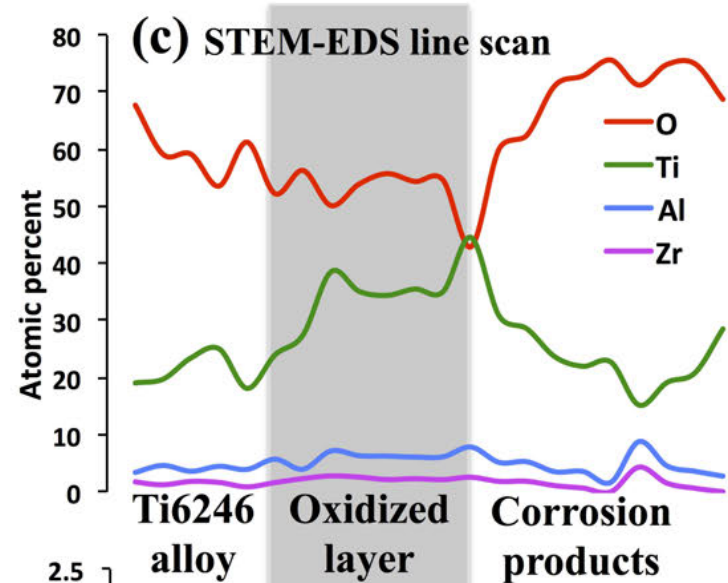
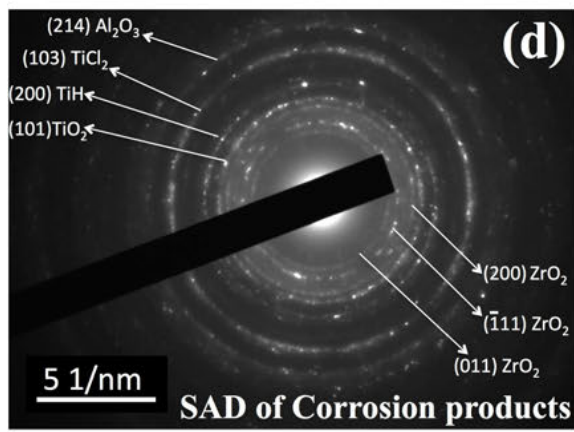 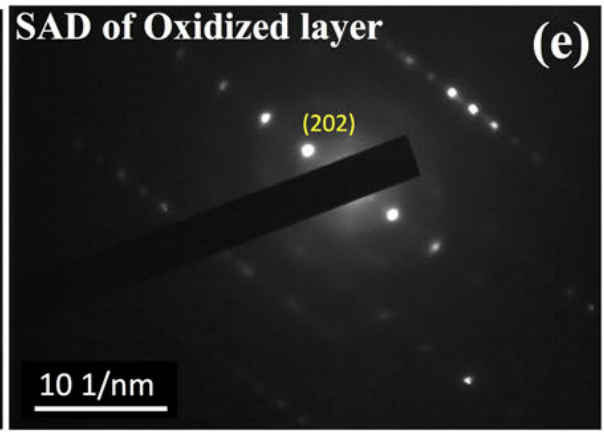 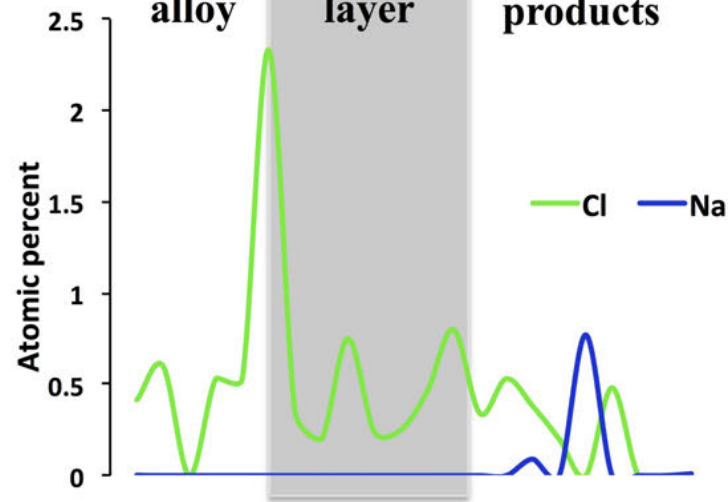

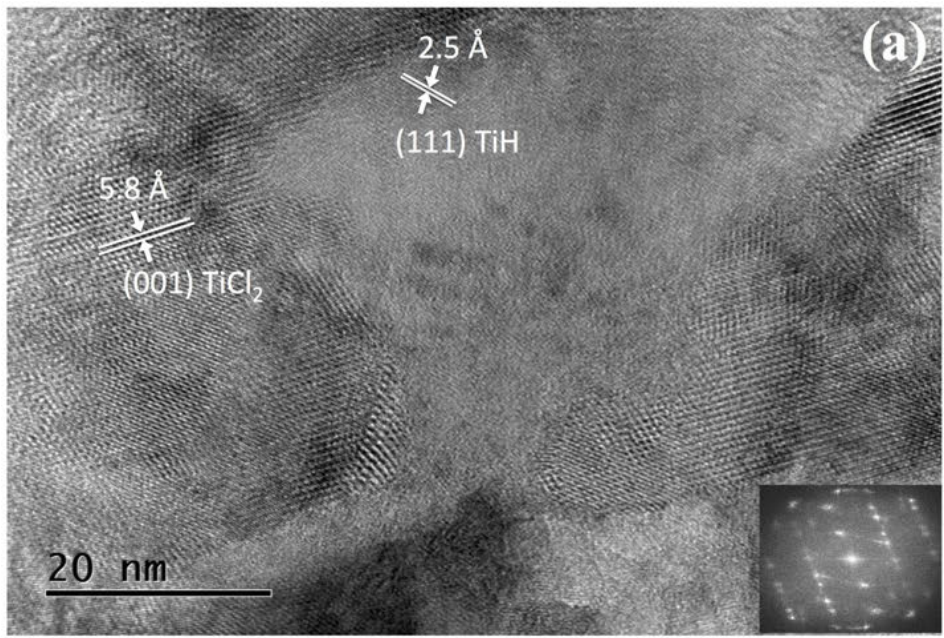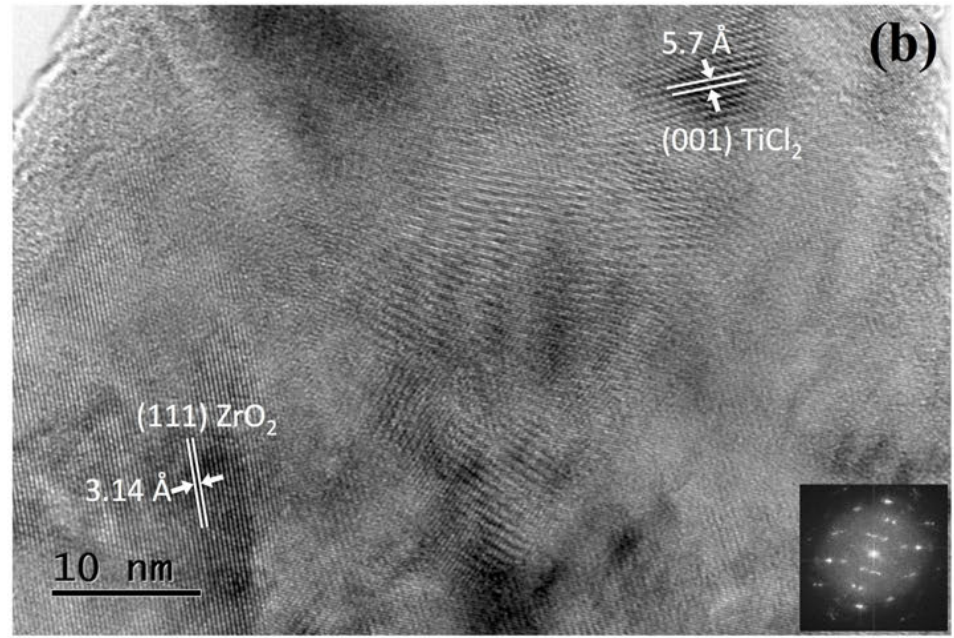

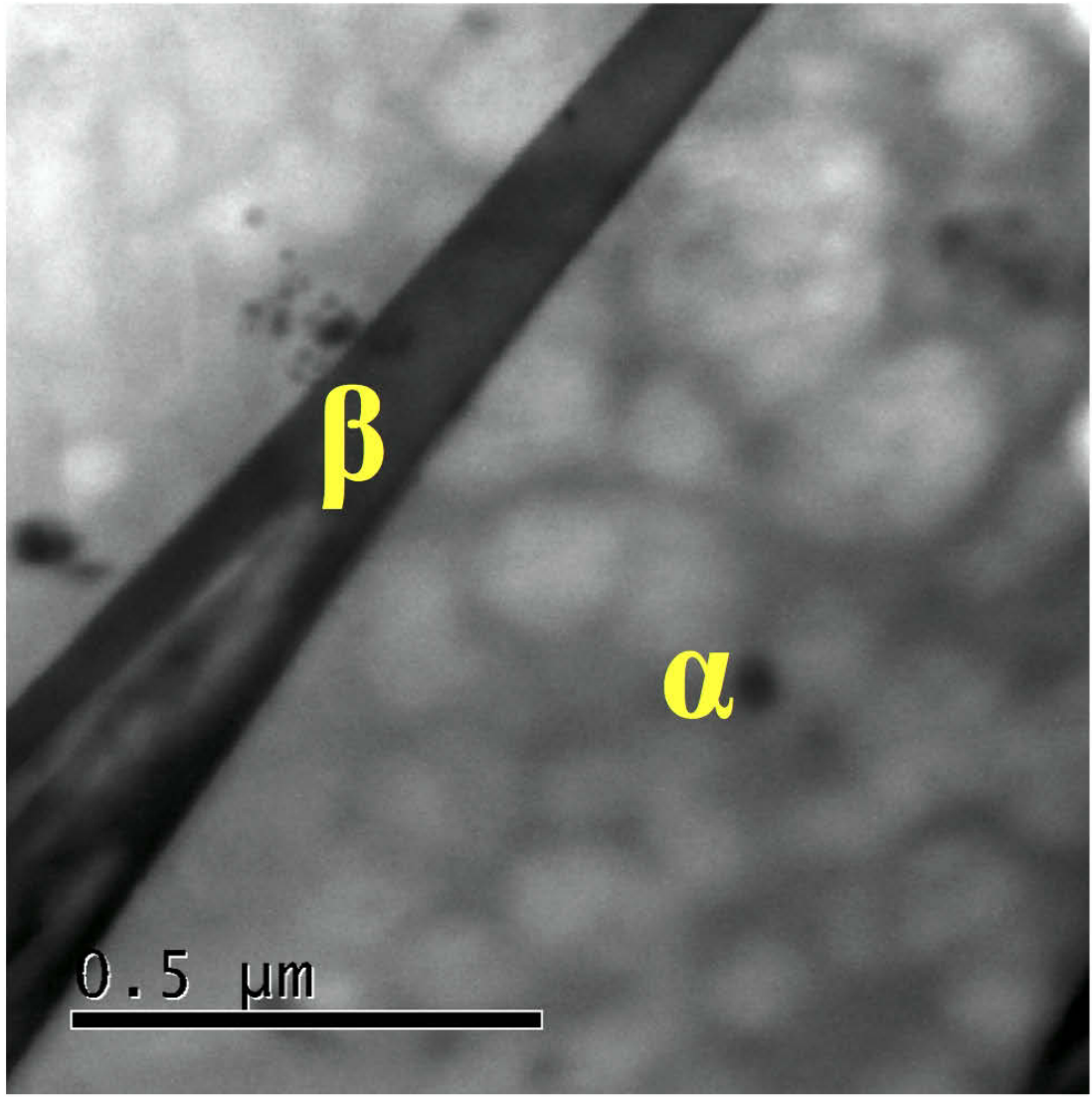

| Element | At.% |
|---|---|
| Na | 2.6 |
| Al | 13.7 |
| Cl | 0.3 |
| Ti | 78.5 |
| Zr | 4.2 |
| Sn | 0.8 |

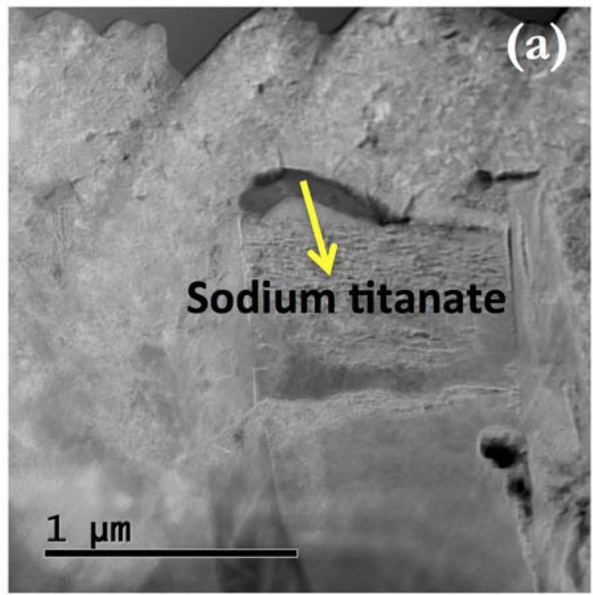 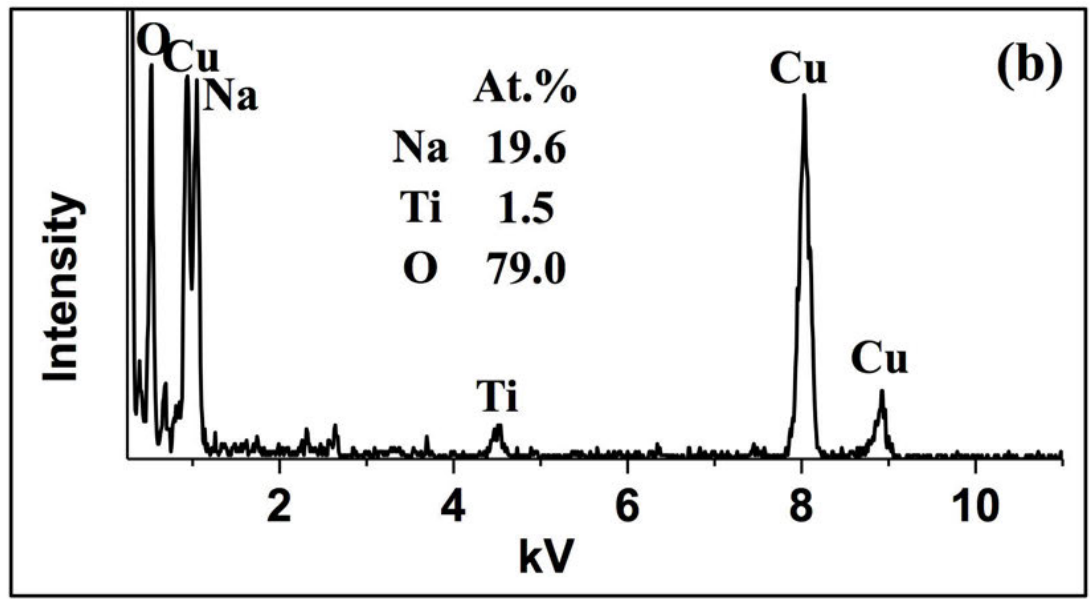
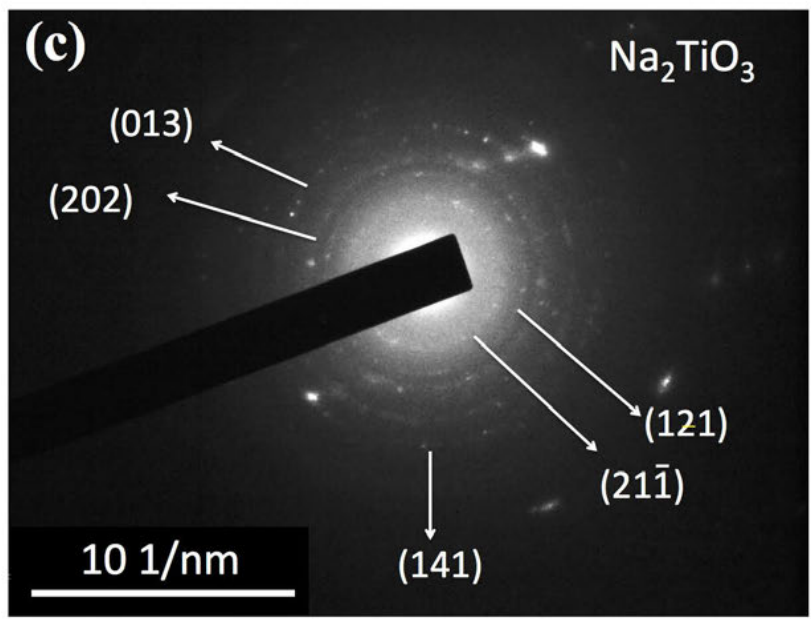 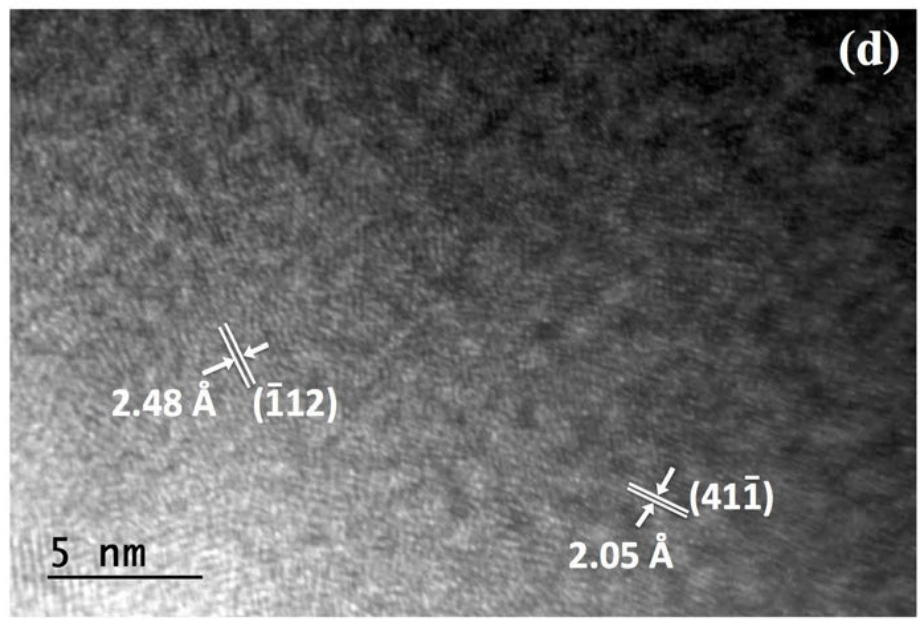

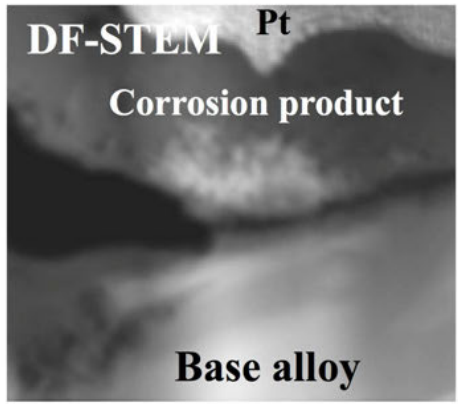
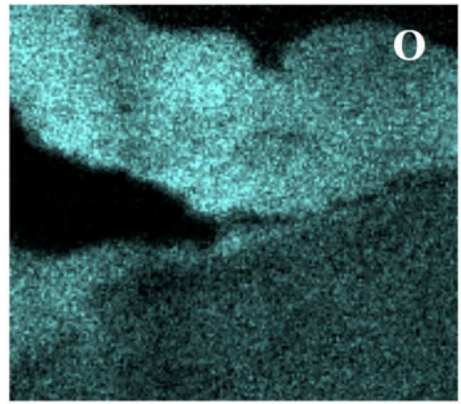
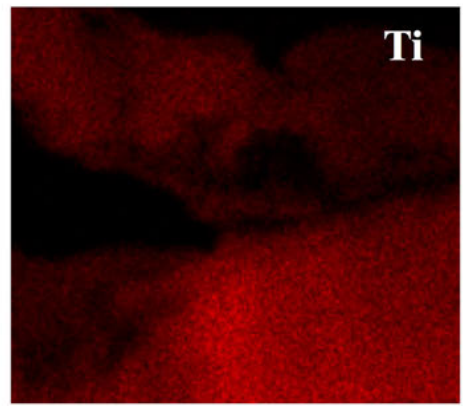
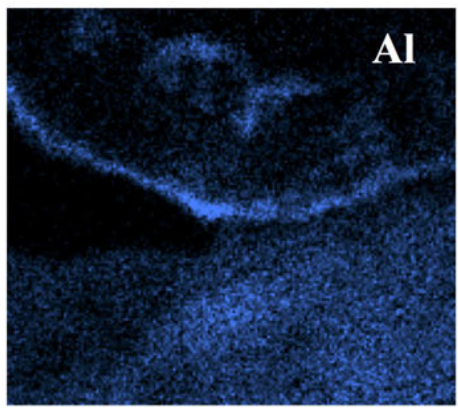
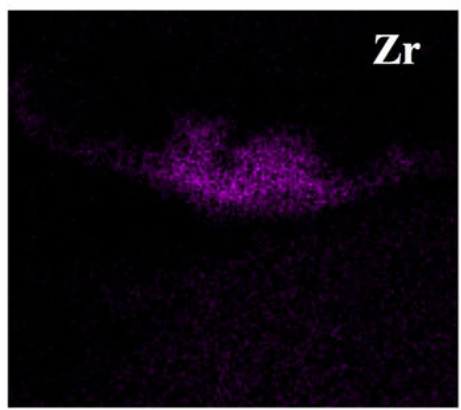
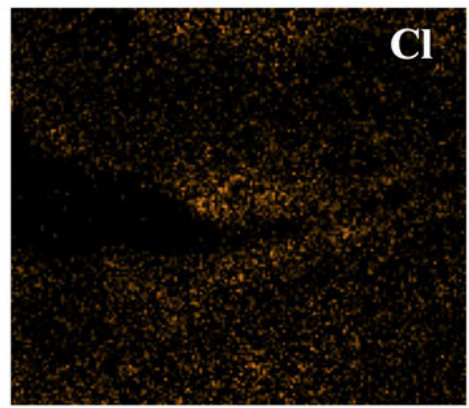
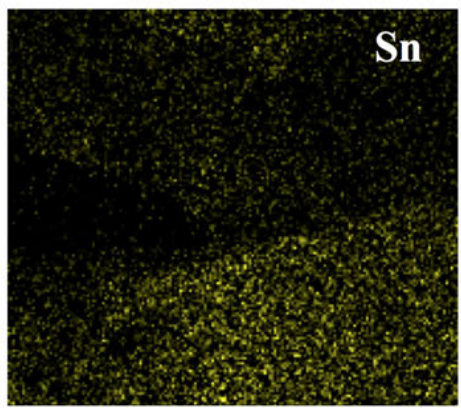
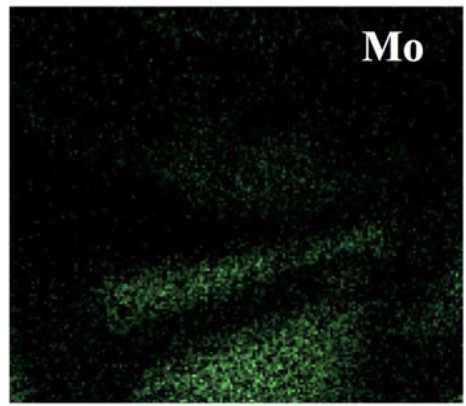

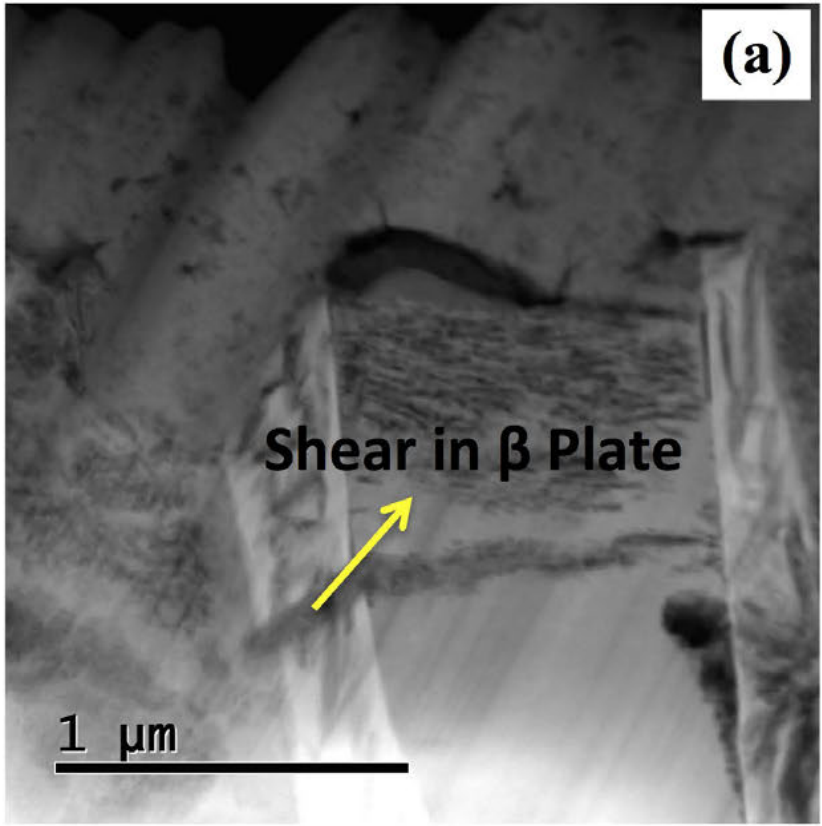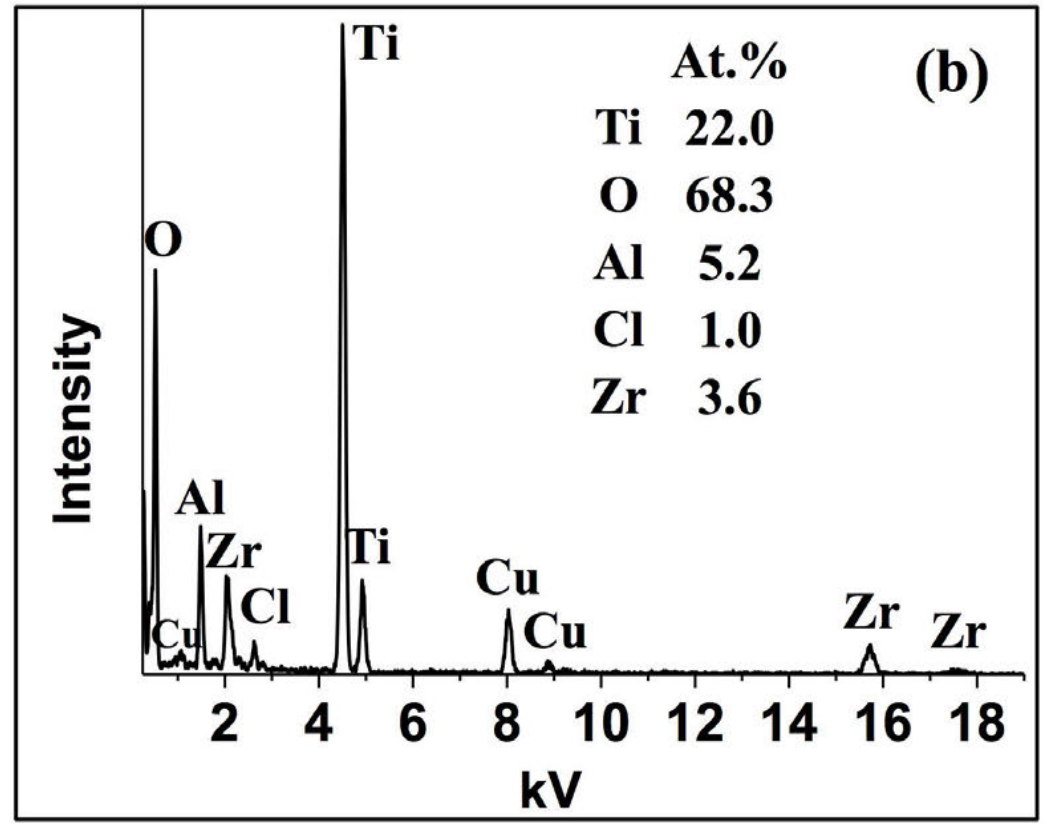

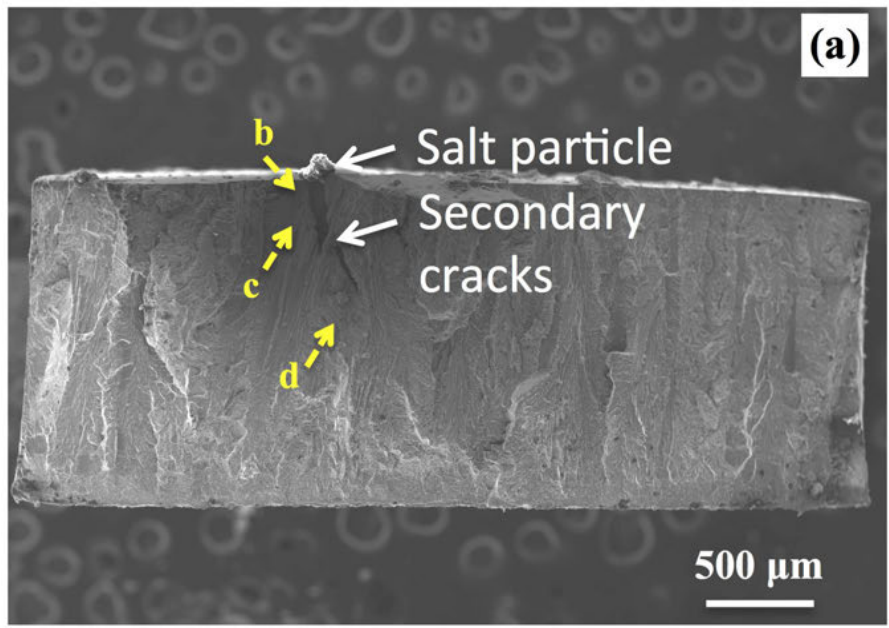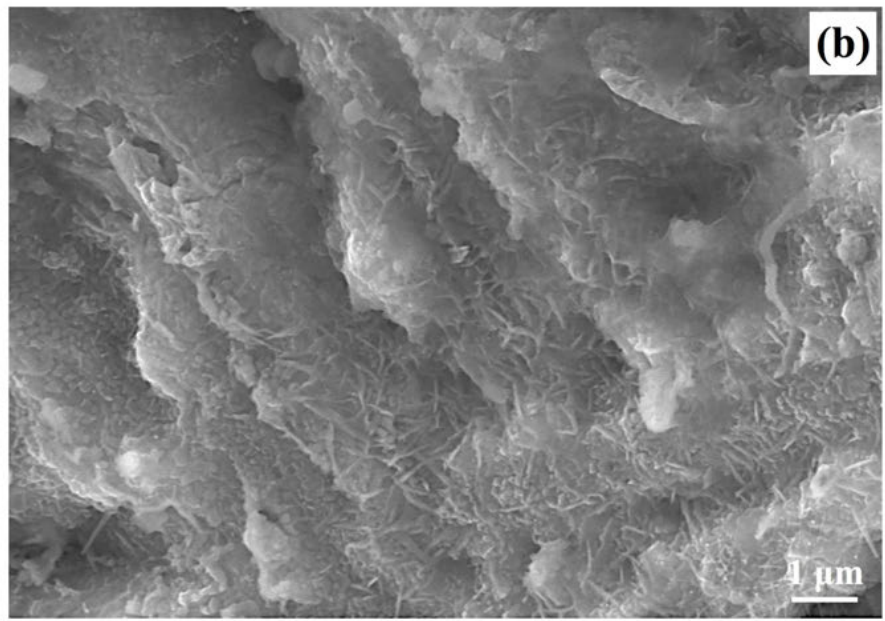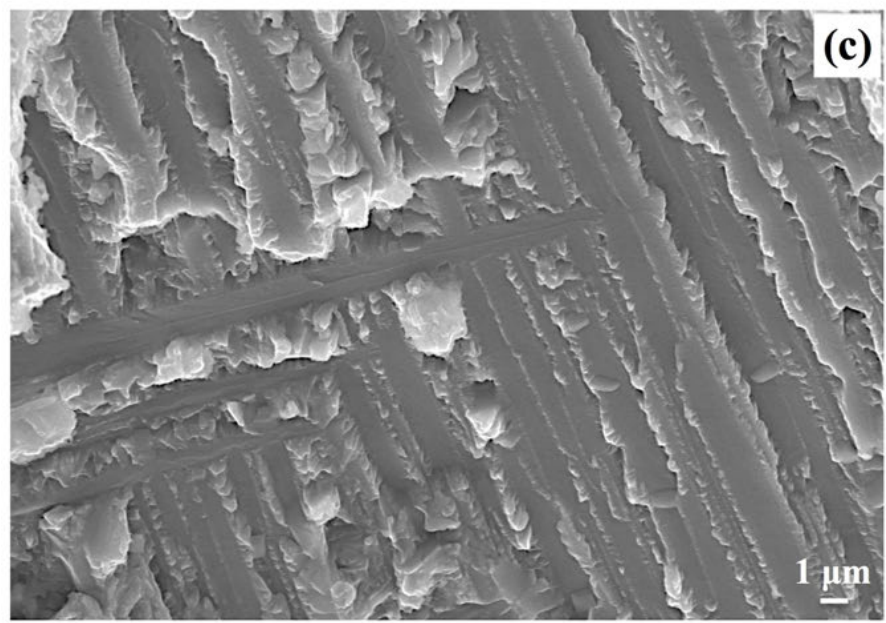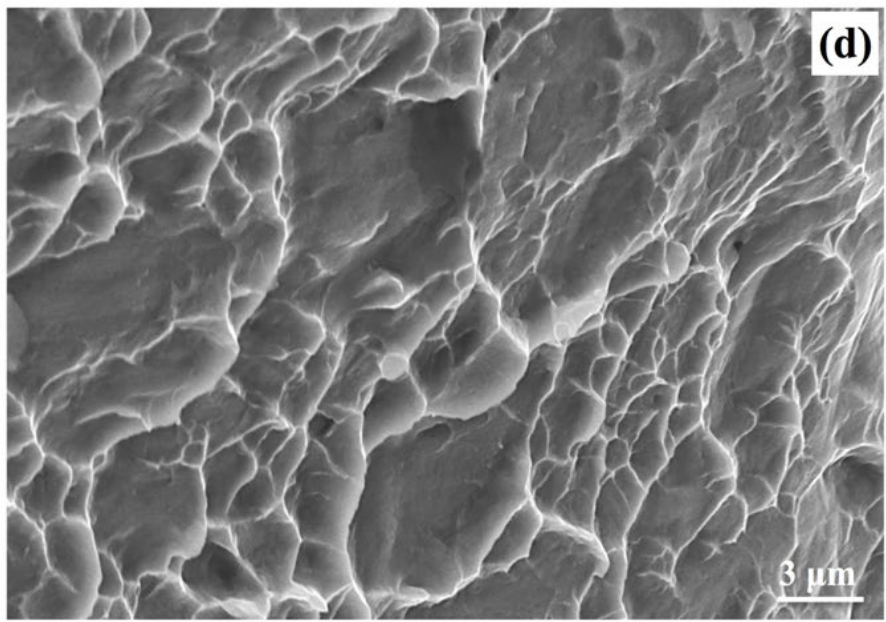

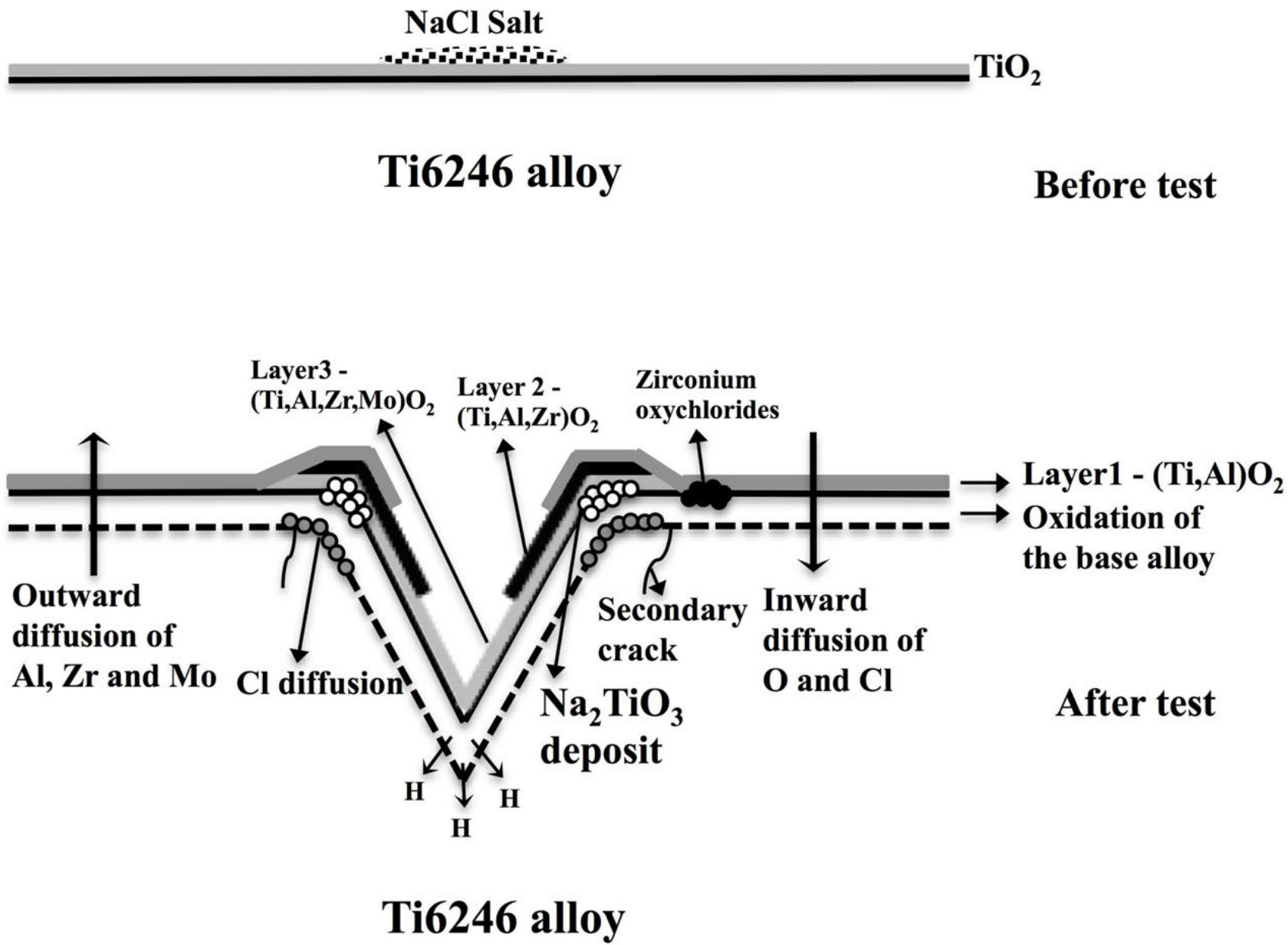